\documentclass[final,5p,twocolumn]{elsarticle}

\usepackage[utf8]{inputenc}

\usepackage[caption=false,font=footnotesize]{subfig}

\begin{document}

\begin{frontmatter}

\title{Photon Detection Efficiency Measurements of the VERITAS Cherenkov
Telescope Photomultipliers after four Years of Operation}

\author[]{Eliza Gazda \corref{mycorrespondingauthor}} \ead{gazdae@my.erau.edu}

\author{Thanh Nguyen} \author{Nepomuk Otte} \author{Gregory Richards}


\cortext[mycorrespondingauthor]{Corresponding author}

\address{School of Physics \& Center for Relativistic Astrophysics, Georgia
Institute of Technology 837 State Street NW, Atlanta, GA 30332-0430, U.S.A.  }

\begin{abstract} The photon detection efficiency of two sets of R10560-100-20
superbialkali photomultiplier tubes from Hamamatsu were measured between 200\,nm
and 750\,nm to quantify a possible degradation of the photocathode sensitivity
after four years of operation in the cameras of the VERITAS Cherenkov
telescopes. A sample of 20 photomultiplier tubes, which was removed from the
telescopes was compared with a sample of 20 spare photomultiplier tubes, which
had been kept in storage. It is found that the average photocathode sensitivity
marginally increased below 300\,nm and dropped by 10\% to 30\% above 500\,nm.
The average photocathode sensitivity folded with the Cherenkov spectrum emitted
by particles in air showers, however, reveals a consistent detection yield of
$18.9\pm0.2$\% and $19.1\pm0.2$\% for the sample removed from the telescope and
the spare sample, respectively.  \end{abstract}


\end{frontmatter}


\section{Introduction}

The Very Energetic Radiation Imaging Telescope Array System (VERITAS)
\cite{Holder2016} is an array of four Cherenkov telescopes that is used to
observe astrophysical objects in gamma rays between 100\,GeV and several
10\,TeV. For an update of the latest results, see \cite{Holder2016}. In the imaging
atmospheric Cherenkov technique (IACT), pioneered by the Whipple Collaboration
\cite{Weekes1989}, large mirror surfaces collect the Cherenkov light that is
emitted by the charged particles in an air shower, which can be initiated by a
gamma ray or a charged cosmic ray in the atmosphere. The Cherenkov light is
projected onto a camera, which images the air shower for an off-line analysis.
In the analysis, based on the shape and intensity of the images, the type of the
air-shower-initiating particle (gamma ray or charged cosmic ray), the energy, and
the arrival direction are reconstructed \cite{Hillas1985}.

The VERITAS reflectors are 12\,m in diameter, and the camera of each telescope
consists of 499 photomultiplier tubes (PMTs). VERITAS has been in operation with all four
telescopes since Fall 2007. In summer 2012 the cameras were upgraded by
replacing the Photonis XP2970 photomultiplier tubes with superbialkali
R10560-100-20 photomultiplier tubes manufactured by Hamamatsu \cite{Otte2011}. The main
reason for the upgrade was the higher photon detection efficiency (PDE) of the
R10560, which resulted in an increased light detection efficiency of the system
and thus better-resolved images, directly impactacting the instrument
performance. The angular resolution improved by 15\%; the energy resolution and
sensitivity both significantly improved, and the lower energy threshold decreased from
100\,GeV to 80\,GeV \cite{Rajotte2014}.

A good knowledge of the PDE is crucial in the event reconstruction because any
offset in the detection yield of the Cherenkov light proportionally
changes the energy scale of the reconstructed gamma rays. To ensure stable
operation, the optical elements of the VERITAS telescopes are, therefore,
regularly monitored in situ. This is done, for example, for the mirrors
\cite{Archambault2013}. The efficiency of the light concentrators in combination
with the photomultipliers is monitored with Cherenkov light from muons
\cite{Vacanti1994}. However, the Cherenkov spectrum from muons peaks at much
lower wavelengths than the spectrum from air showers due to the lack of
atmospheric scattering and absorption.\footnote{The Cherenkov light collected
from muons originates closer to the detector.} For a better test of the photon
detection efficiency, it is necessary to remove the photomultiplier tubes from
the telescope and measure them in the laboratory.

In order to check whether any degradation of the sensitivity of the photocathode
has occurred after the Hamamatsu PMTs were installed in 2012, we performed
wavelength-dependent PDE measurements between 200\,nm and 750\,nm on a subset of
the 1,996 photomultiplier tubes currently installed. For that purpose 5 PMTs were
temporarily removed from each of the four VERITAS cameras in summer 2016, and
their PDEs were measured in the lab at Georgia Tech. The PMTs were reinstalled before
the end of the summer. The results are compared to the PDEs of 20 spare PMTs,
which were measured at the same time as the PMTs pulled from the telescopes. The
spare PMTs come from the same production batches as the PMTs used in the
telescopes and have been stored in a cool, dry, and dark location at the VERITAS
site and never been used in a telescope.

\section{The R10560 Photomultiplier Tube}

\begin{figure}[!tb] \centering
\includegraphics[width=\columnwidth]{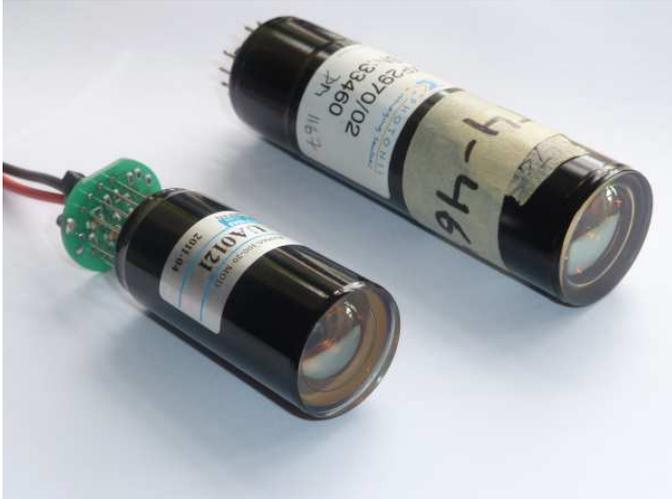} \caption{Photograph
of the Photonis XP 2970 (right) that was replaced with the Hamamatsu
R10560-100-20 (left) in 2012.} \label{PMTPic} \end{figure}

The R10560-100-20 MOD is shown in Figure \ref{PMTPic} together with its
predecessor used in VERITAS, the Photonis XP 2970. The R10560 is a one-inch PMT
with a UV-transparent entrance window and a superbialkali photocathode with a
peak efficiency of $\sim33\%$ at 300\,nm. The R10560 has a linear focused dynode
structure with eight stages that is biased in a 4, 1, 1, 1, 1, 1, 1, 1, 1
configuration. The voltage divider consists of a passive divider chain between
the cathode and dynode 5. Between dynode 6 and the anode, the bias is stabilized by
an active divider circuit. 

Aging of the photomultiplier tubes is a concern in Cherenkov telescope
applications with respect to the PMT gain and photocathode sensitivity. VERITAS
operates during nights with less than 70\% moon illumination. Due primarily to
stars, zodiacal light, and anthropogenic light sources, the R10560 typically detects an 
ambient background photon rate (also called night sky
background) of $70\cdot10^6$ to $150\cdot10^6$ counts/second. With a
multiplication gain of the PMTs of about 200,000, the anode current is typically a
few microamperes. In bright moonlight conditions, the bias of the PMTs is
reduced to keep the current below 10\,$\mu$A. After four years of operation, the
last dynode of a PMT has thus typically collected a charge of $\le$100\,C. To
compensate for the loss in gain due to the bombardment of the last dynodes with
electrons, the PMT bias is continuously readjusted to keep the gain within 5\% of
its nominal value. 

Aging of the photocathode can, for example, occur due to positively ionized
residual gas molecules inside the PMT that are accelerated toward the
photocathode and damage it upon impact. Another possibility is aging due to high
ambient temperatures. During summer months the daytime temperatures at the
VERITAS site are regularly above 40$^{\circ}$C and even higher inside the
camera.  \section{PDE Measurement Procedure}

\begin{figure}[!tb] \centering
\includegraphics[width=\columnwidth]{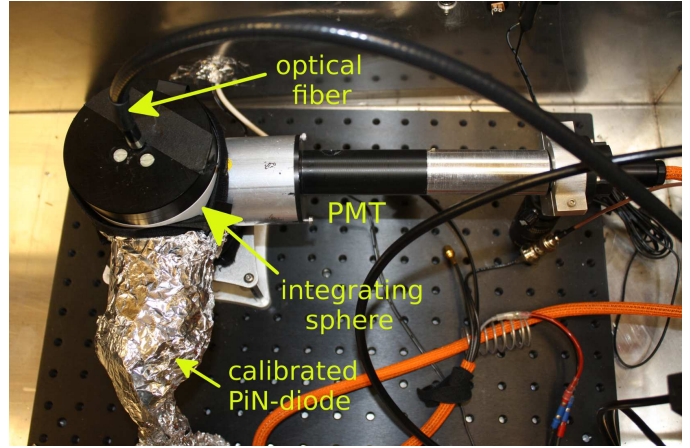} \caption{Setup to
measure the PDE of the PMTs at four wavelengths.} \label{PDEMeasPic}
\end{figure}

\begin{figure}[!tb] \centering
\includegraphics[width=\columnwidth]{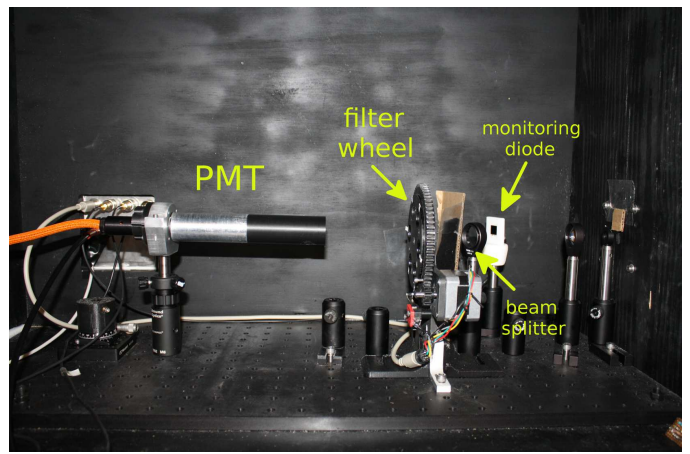}
\caption{Setup to measure the spectral response of the PMTs. Shown here is the
configuration with the PMT installed. The response of a calibrated
\emph{Si}-diode is measured by replacing the PMT with the \emph{Si}-diode.}
\label{SpectrPic} \end{figure}

The PDE measurement procedure consists of three steps and is explained in detail
in \cite{Otte2016}. In the first step the PDE is measured at four different
wavelengths (400\,nm, 452\,nm, 500\,nm, and 589\,nm). In the second step the
relative spectral response is measured between 200\,nm and 750\,nm, which is the
wavelength range where the efficiency of the PMTs is non zero. In the last step
the relative spectral response is scaled to the four PDE measurements to arrive
at an absolute PDE measurement between 200\,nm and 750\,nm. Figure
\ref{PDEMeasPic} shows a picture of a PMT mounted in the PDE setup, and Figure
\ref{SpectrPic} shows a picture of a PMT in the spectral response setup.

\begin{figure}[!tb] \centering
\includegraphics[width=\columnwidth]{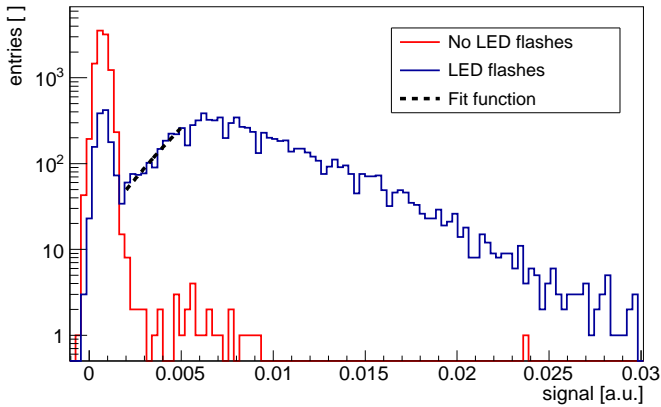} \caption{Pulse height
distribution of the PMT signals when the PMT is not flashed with a light source
(red) and when it is flashed (blue). The dashed black line shows the
fit to the single photoelectron distribution, which is extrapolated into the
pedestal and subtracted from it.} \label{PMTPHD} \end{figure}

The only difference with respect to the procedure as described in
\cite{Otte2016} is how the number of detected photoelectrons is extracted from
the PMT data. The setup was originally devised to measure the PDE of silicon
photomultipliers, which exhibit an excellent separation of single photoelectron
signals from noise. In PMTs the single photoelectron signals partially overlap
with noise because of the statistical nature of the multiplication process in
the dynode structure of the PMT. Figure \ref{PMTPHD} shows an example of the
pulse height distribution recorded with and without illumination by a flashing
light source in the PDE setup. While for SiPMs it is sufficient to define a
signal level above which all signals can be considered to be from detected
photons and below which from noise, the same approach applied to PMT signals
would result in considerable systematic uncertainties.

In order to alleviate the problem, we fit the portion of the single
photoelectron distribution between the pedestal and the peak of the single
photoelectron distribution with a Gaussian (dashed line in Figure \ref{PMTPHD})
and extrapolate the fit into the pedestal. We then subtract the extrapolation
from the pedestal distribution, which enables us to extract the number of flashes during
which the PMT did not detect a photon. The number of non-detections is then used
to calculate the average number of photoelectrons detected during a flash as
described by Equation 2 in \cite{Otte2016}.

The PMTs are biased at 1200\,V in the PDE measurements, which is higher than the
typical operating bias of most tubes (600\,V - 1300\,V), to produce a better
separation between noise and single photoelectron signals. The PDE does not change 
when the bias is increased even higher to 1400\,V.

The overlap between the pedestal distribution and the single photoelectron
distribution, however, remains a major source for systematic uncertainties
because the precise shape of the single photoelectron distribution is unknown.
We estimate that the total systematic uncertainty of the PDE measurement is 6\%
based on the uncertainties quoted in \cite{Otte2016} and a 5\% uncertainty on
the previously described method to extract the number of photoelectrons from the
pulse height distribution.

In the PDE measurement a random, 1\,mm diameter spot approximately in the center
of the photocathode is illuminated. In the relative spectral response
measurement, the light beam illuminates an area of the photocathode with a
diameter of about 10\,mm.

\section{Results}

\begin{figure}[!ht] \subfloat[PMTs in-use in the VERITAS
telescopes]{\includegraphics[width=\columnwidth]{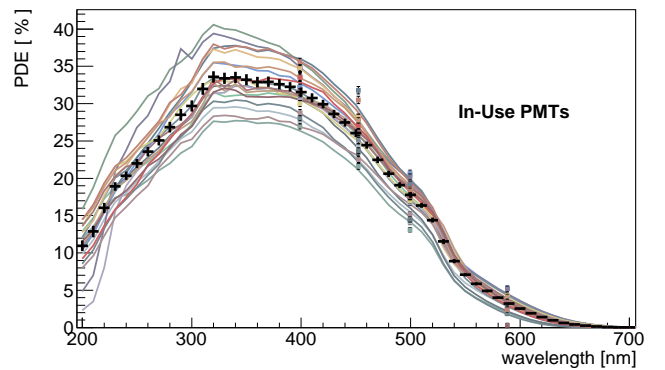}\label{PMThot}}\\
\subfloat[Spare
PMTs]{\includegraphics[width=\columnwidth]{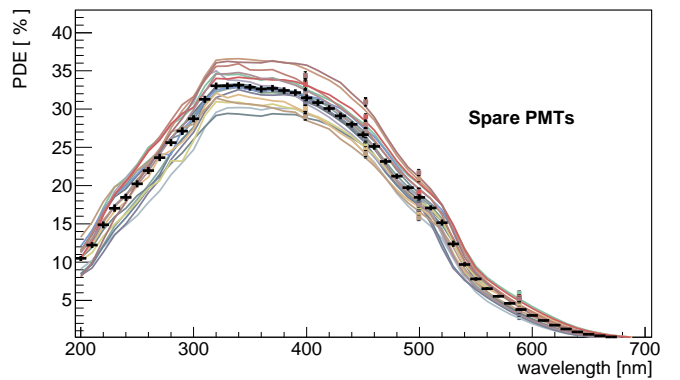}\label{PMTSpare}}
\caption{PDE measurements of the 20 PMTs removed from the VERITAS telescopes and
20 spare PMTs between 200\,nm and 750\,nm.\label{PDEvsBias}} \end{figure}

\begin{figure}[!tb] \centering
\includegraphics[width=\columnwidth]{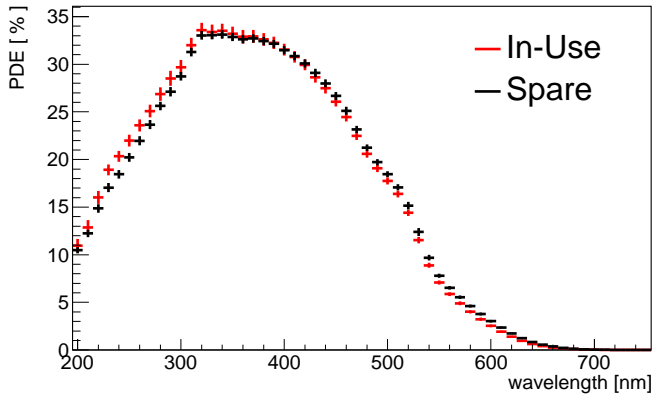} \caption{Average PDE of
both sets of PMTs.} \label{PDEAverage} \end{figure}

Figures \ref{PMThot} and \ref{PMTSpare} show the absolute PDE of the two sets of
PMTs between 200\,nm and 750\,nm in steps of 10\,nm. Depicted by the thicker
data points at 400\,nm, 452\,nm, 500\,nm, and 589\,nm are the PDE measurements
to which the relative spectral response has been scaled. The black data points
are the average values of all twenty PMTs. The vertical error bars show the
error on the mean values, and the horizontal error bars represent the 10\,nm steps. Figure
\ref{PDEAverage} shows the two average PDE curves. It is
evident from these figures that a mismatch exists between the two average curves,
and that the scatter in the PDE curves is larger for the PMTs removed from the
VERITAS telescopes (hereafter \emph{in-use} PMTs) than for the \emph{spare}
PMTs.

\begin{figure}[!tb] \centering
\includegraphics[width=\columnwidth]{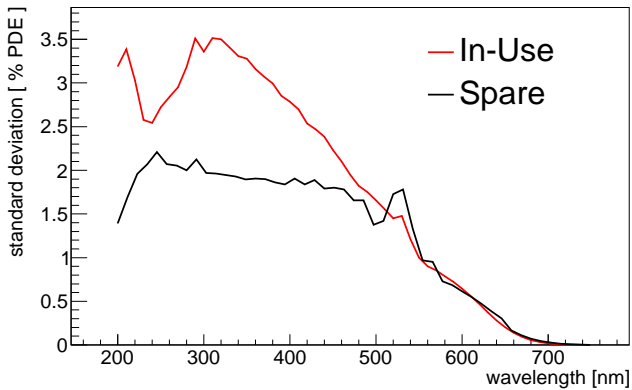} \caption{Standard
deviation of the PDEs for both sets of PMTs.} \label{Variance} \end{figure}

The scatter of the PDE curves is quantified by calculating the root mean square
(RMS) of the PDE values for each set in steps of 10\,nm. The results are shown
in Figure \ref{Variance}. Above 500\,nm both sets of PMTs display approximately the same RMS.
Below 500\,nm, however, the RMS of the \emph{in-use} PMTs is consistently above
that of the \emph{spare} PMTs and peaks at 300\,nm with a 50\% higher RMS than
the RMS for the \emph{spare} PMTs. The higher RMS does not impact VERITAS
operations because the 3\% PDE RMS at 300\,nm
corresponds to only 10\% of the peak PDE at 300\,nm.

\begin{figure}[!tb] \centering
\includegraphics[width=\columnwidth]{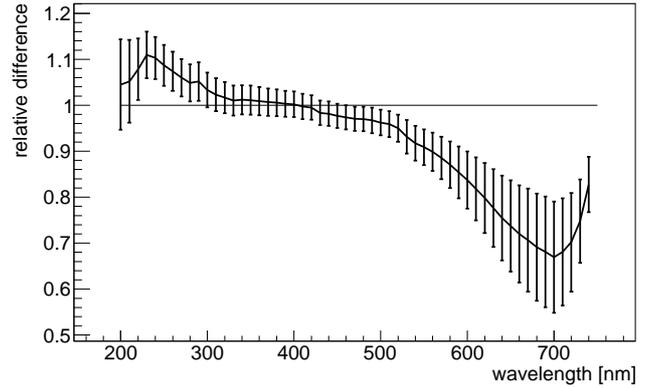}
\caption{Relative difference of the average PDE of the \emph{in-use} set of PMTs
relative to the average PDE of the \emph{spare} set of PMTs.} \label{PDEresid}
\end{figure} 

The difference in the average PDE values between the two samples is
quantified by calculating the residual values of the average PDE of the
\emph{in-use} PMTs relative to the average PDE of the \emph{spare} PMTs; see
Figure \ref{PDEresid}. While there is evidence of an increase in the PDE below
300\,nm, the PDE of the \emph{in-use} PMTs is lower than the PDE of the
\emph{spare} PMTs above 500\,nm. The difference reaches its maximum at 700\,nm
where the \emph{in-use} PMTs show on average a 30\% lower efficiency.  It
should be noted that the absolute PDE drops below 7\% at 550\,nm and is below
2\% at 600\,nm and above. Thus, even a relative 30\% drop does not significantly
affect VERITAS operation as we will discuss next.

\begin{figure}[!tb] \centering
\includegraphics[width=\columnwidth]{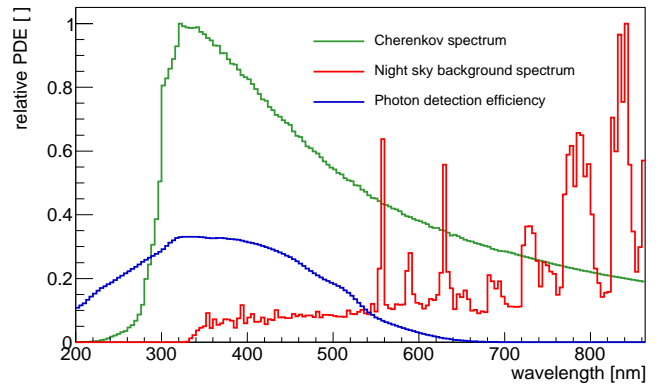} \caption{Simulated
Cherenkov spectrum in the focal plane of a VERITAS camera, the night sky
background spectrum, and the average PDE of the PMTs removed from the telescopes
and tested here. The spectra are normalized to a peak intensity of one.}
\label{FoM} \end{figure} The PMTs detect Cherenkov light from air showers. The
Cherenkov intensity is proportional to $1/\lambda^2$ where $\lambda$ is the
photon wavelength. Below 340\,nm the Cherenkov emission is scattered and
absorbed in the atmosphere. Figure \ref{FoM} shows a simulated Cherenkov
spectrum in the focal plane of a VERITAS camera. The full width at half maximum
of the spectrum spans the range from 300\,nm to 520\,nm, which is about the
wavelength range over which a photon detector should be sensitive to yield
maximal detection efficiency for the Cherenkov light. Above 550\,nm the
intensity of the night sky background becomes dominant, and it is therefore desirable
that the sensitivity of the photon detector drops to zero. Figure \ref{FoM}
shows a measured spectrum of the night sky background \cite{Benn1999}. Also
shown in the figure is the average PDE of the \emph{in-use} PMTs measured in
this campaign.

The detection yield of Cherenkov light is the figure of merit that can be used
to judge if the measured changes in the PMT sensitivity have a significant
impact on VERITAS operation. To find out, we multiplied the simulated Cherenkov
spectrum in Figure \ref{FoM} by the average PDE and integrated the product
from 200\,nm to 750\,nm. Normalizing to the integral of the Cherenkov spectrum
over the same wavelength range gives the fraction of the light detected by the
PMTs. We obtain $18.9\pm0.2$\% for the yield of the \emph{in-use} PMTs and
$19.1\%\pm0.2$\% for the yield of the \emph{spare} PMTs. Thus both sets have the
same detection yield to air-shower Cherenkov light.

\section{Discussion}

We evaluated two sets of Hamamatsu R10560 PMTs for the purpose of investigating
a possible degradation of the photocathode sensitivity after four years of
operation in the VERITAS Cherenkov telescopes. The previously used PMTs, the
Photonis XP2970, showed aging of the spectral sensitivity, and it was not clear if
a similar effect would be seen in the new PMTs. Some Photonis PMTs even showed a
complete loss of their sensitivity.  A degradation of the UV sensitivity was
also observed in the Hamamatsu R\ 1398 PMTs that were used by the Whipple
Collaboration \cite{Daniel2016, 2007APh....28..182K}.

For the measurements one set of 20 PMTs was temporarily removed from the VERITAS
telescopes and reinstalled after the measurements had been completed. The second
batch of 20 PMTs consists of unused spare PMTs, which have never been installed
in the telescopes.

In this study, we have found that the spectral sensitivity of the PMTs significantly 
degraded above 500\,nm, and
that there is evidence of a slight improvement of the sensitivity below 300\,nm.
However, both changes do not affect the detection yield of Cherenkov light, which we
have shown by calculating the yield for the two sets of PMTs. 

In conclusion, the Hamamatsu R10560-100-20 superbialkali PMT did not
degrade in performance after four years of operation during which each PMT anode
accumulated an estimated charge of approximately 100\,C \cite{Otte2011} and
experienced extreme temperature variations with peak temperatures of more than
40$^{\circ}$C.

\section*{Acknowledgments}

We are grateful for the work done by the staff at the VERITAS site to send us
PMTs for testing.  Her research was funded with grants from the Center on
Materials and Devices for Information Technology Research (CMDITR), the NSF
Science and Technology Center No. DMR 0120967, and the National Science
Foundation under grant no. PHYS-1505228. E.\ G.\ would like to thank her advisor
N.\ O.\ and all of the REU organizers at Georgia Tech for their knowledge and
support received during this summer project.

\section*{References} \bibliography{bibliography}

\end{document}